# A 660 – 2000 nm Multibranch Laser Frequency Comb based on an Erbium-Doped Fiber Laser

## Gabriel Ycas,[*1,2] Steve Osterman,[3] and Scott A. Diddams[2,4]

[1] Department of Physics, University of Colorado, Boulder, CO
[2] National Institute of Standards and Technology, 325 Broadway, Boulder, CO
[3] University of Colorado, Center for Astrophysics and Space Astronomy, Boulder, CO
[4]sdiddams@boulder.nist.gov
*Corresponding author: ycasg@colorado.edu

We present a multibranch laser frequency comb based upon a 250 MHz mode-locked erbium-doped fiber laser that spans more than 300 terahertz in bandwidth, from 660 nm to 2000 nm. Light from a mode-locked Er:fiber laser is amplified and then broadened in highly-nonlinear fiber to produce substantial power at ~1050 nm. This light is subsequently amplified in Yb:fiber to produce 1.2 nJ, 73 fs pulses at 1040 nm. Extension of the frequency comb into the visible is achieved by supercontinuum generation from the 1040 nm light. Comb coherence is verified with cascaded f-2f interferometry and comparison to a frequency stabilized laser.
 OCIS Codes: 190.7110, 060.2320

The Er:fiber-based laser frequency comb is a technologically mature tool for precision metrology with benefits of long-term reliability and relatively low cost. However, in some applications, such as direct frequency comb spectroscopy [1, 2] and the calibration of high-precision astronomical spectrographs [3], it would be valuable to continuously extend the wavelength coverage of the Er:fiber-based laser frequency comb (LFC) below 1 μm and into the visible. Earlier efforts have accomplished this by making use of second harmonic generation and nonlinear broadening [4], off-resonant quasi-phase matched interactions in waveguides [5], or a combination of large mode-area fibers and cascaded nonlinear fibers [6]. Here, we present a technique for generating a visible light frequency comb from a mode-locked erbium laser that is self-referenced and frequency-stabilized using established techniques. In nonlinear fiber we shift significant pulse energy to the 1 μm region, where it is amplified with a core-pumped Yb:fiber amplifier [7, 8, 9, 10, 11]. The amplifier output is then compressed to provide a 73 fs pulse, which is spectrally broadened to below 650 nm using microstructured fiber. Significantly, we verify that the coherence of the original Er:fiber source is transfered to the amplified 1040 nm pulses, and that with subsequent nonlinear broadening the system provides a multibranch LFC with nearly two octaves of bandwidth from 660 nm to beyond 2000 nm.

Intense ultrashort pulses at 1.04 μm are generated from a 250 MHz mode-locked Er:fiber laser [12] using a series of amplifiers and nonlinear fibers, shown in Fig. 1. One third (35 mW) of the light produced by the Er:fiber laser is amplified in a core-pumped Er:fiber amplifier to an average power of 450 mW. Dispersion management and nonlinear pulse-shortening are achieved by carefully trimming the length of standard anomalous-dispersion single-mode fiber (SMF) between the laser and the gain fiber and by making use of a normal-dispersion Er:doped fiber [13] (nLight Er80 4/125). After recompressing the amplified pulses in SMF, a pulse duration of ~ 70 fs at 1580 nm is achieved as measured using second-harmonic generation frequency-resolved optical gating (SHG-FROG) [14].

The erbium amplifier output is fusion spliced to a 5 cm long piece of solid-core highly nonlinear fiber (HNLF) [15] with dispersion of 7.7 (ps/nm)/km at 1600 nm, which is in turn coupled out by splicing to a length of SMF. This generates a supercontinuum with ~ 8 % of the 240 mW power coupled into the SMF falling between 1000 nm and 1100 nm. Within that range, the placement of the spectral peak can be refined by tuning the polarization state of light entering the Er:fiber amplifier and the HNLF. For example, to allow amplification in Yb:fiber at 1030 (1050) nm, the peak can be centered at 1027 (1065) nm, with 14 (11.5) mW of power in a 70 (100) nm bandwidth. The HNLF output is then coupled into a core-pumped ytterbium fiber amplifier that provides an amplified average power of 400 mW at a pump power of ~ 1 W. For environmental stability, the entire fully-spliced amplifier and HNLF apparatus is placed inside a small box.

Using a volume phase holographic single grating compressor [16], the amplified 1040 nm pulses are compressed with 75% power efficiency to 73 fs duration, as measured by SHG-FROG (Fig. 2). To generate a spectrum extending into the visible, these pulses are coupled into a 0.5 m microstructured nonlinear fiber with a zero-dispersion wavelength of 945 nm. Accounting for losses in the grating compressor and fiber coupling, about 110 mW of power was launched into the microstructured fiber. This generates an octave-spanning spectrum from 660 nm to 1340 nm (Fig. 1).

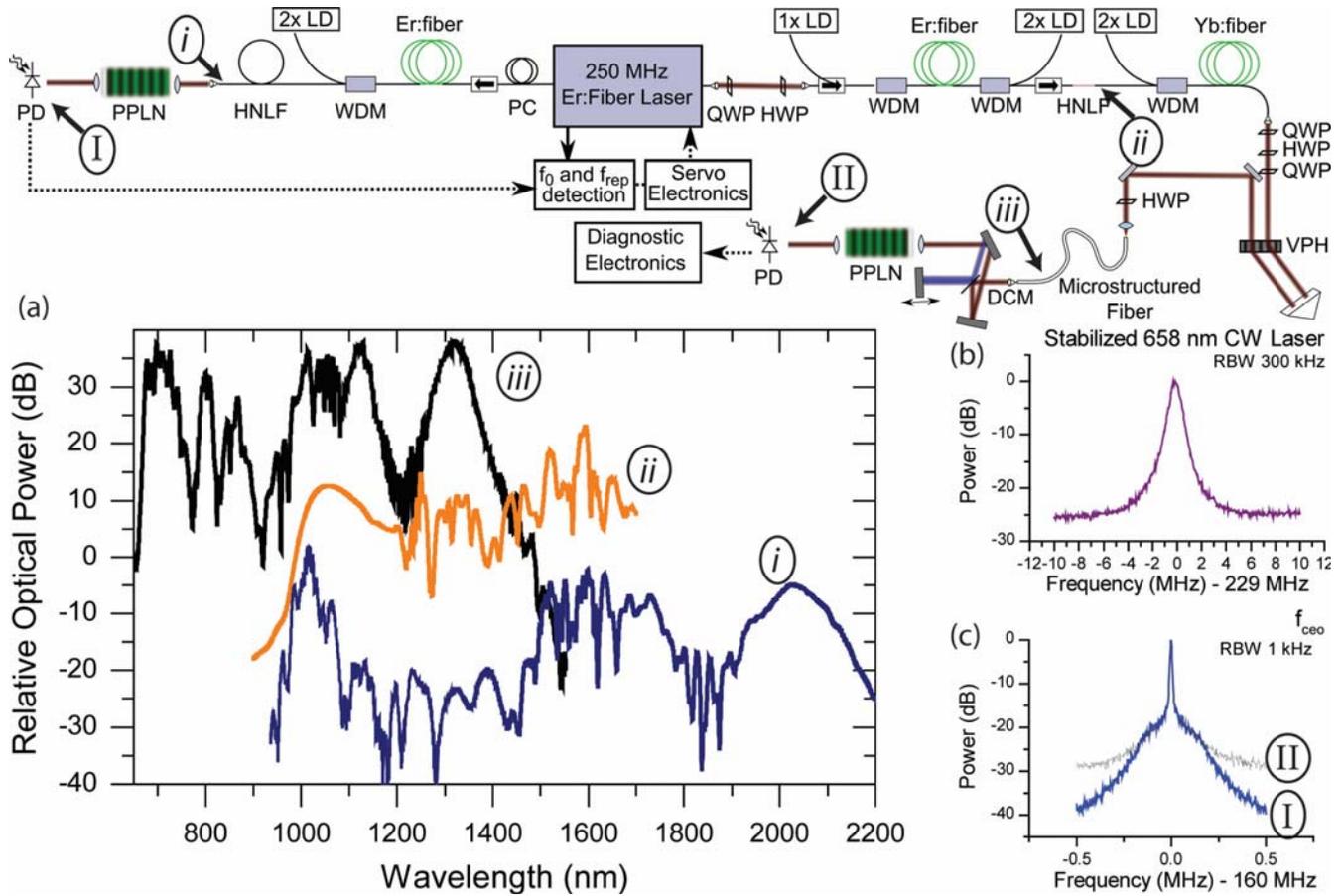

Fig. 1. Top: Schematic of apparatus used to generate and characterize 660 nm—1400 nm spectrum. LD: 600 mW, 976 nm laser diode; WDM: wavelength division multiplexer; Q(H)WP: quarter (half) waveplate; PC: polarization controller; VPH: volume phase holographic grating (Kaiser Optical Systems); DCM: dichroic mirror; PPLN: periodically-poled lithium niobate; PD: photodiode; RFSA: radio-frequency spectrum analyzer. (a) Optical spectra taken at locations (i), (ii), and (iii) in the system, offset for clarity. (b) Heterodyne beat note of the LFC with a 657 nm cavity-stabilized laser, (c) the $f_{CEO}$ beat from (I) 1 μm to 2 μm and (II) 660 nm to 1340 nm; all acquired with LFC stabilized.

In order to verify that the light produced by the chain of amplifiers and nonlinear fibers retains the coherence of the mode-locked laser, a series of heterodyne measurements were conducted. First, a second output from the mode-locked Er-laser was used to generate an octave-spanning spectrum from 1 μm to 2 μm, which enabled frequency detection of the carrier-envelope offset frequency ($f_{CEO}$) with a standard f-2f interferometer (free-running SNR of 35 dB in 300 kHz BW) and stabilization via feedback to the laser's pump current. The laser repetition rate ($f_{rep}$) was also stabilized relative to a Rb clock or hydrogen maser. Subsequently, utilizing the octave-spanning spectrum from the microstructured fiber, a second f-2f interferometer between 660 nm and 1320 nm was constructed to detect an out-of-loop copy of $f_{CEO}$ (free-running SNR of 25 dB in 300 kHz BW). We then used a high-resolution (Λ) frequency counter [17] to characterize the instability of the in-loop (1 μm to 2 μm) and out-of-loop $f_{CEO}$ beats. The Allan deviations computed from the time series of 1 s counter readings are shown in Fig. 3. As seen, the instability of the in-loop $f_{CEO}$ is counter-limited near $2 \times 10^{-18}$ at 1 s of averaging time, while the instability of the out-of-loop $f_{CEO}$ is only 640 mHz deviation (fractionally $1.4 \times 10^{-15}$) at 1 s averaging time.

We also measured the phase noise of the two copies of $f_{CEO}$ (Fig. 4). Above Fourier frequencies of approximately 400 Hz, the two sets of data are nearly identical. Quantitatively, the integrated phase noise from 1 MHz to 1 Hz is seen to increase from 0.61 (in-loop) to 0.78 radians (out-of-loop.) At low frequency, the phase noise of the out-of-loop $f_{CEO}$ increases significantly – consistent with the counter data of Fig. 3, and is most likely a result of thermal and acoustic perturbations. Notably, there is only a 20% reduction in the relative coherence between the two $f_{CEO}$ beats. This excess noise could be removed with a low-bandwidth servo [11].

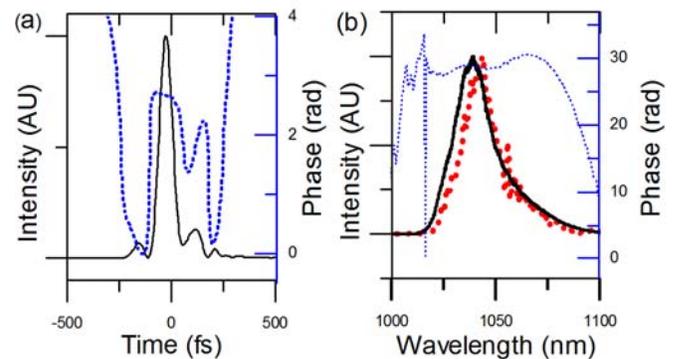

Fig. 2. (a) Amplitude and phase of the amplified 1040 nm pulse after compression, retrieved by SHG-FROG. The FWHM pulse duration is 73 fs and the average power is 300 mW. (b) Optical spectrum measured with linear spectrometer (solid line) and retrieved with FROG (dotted).

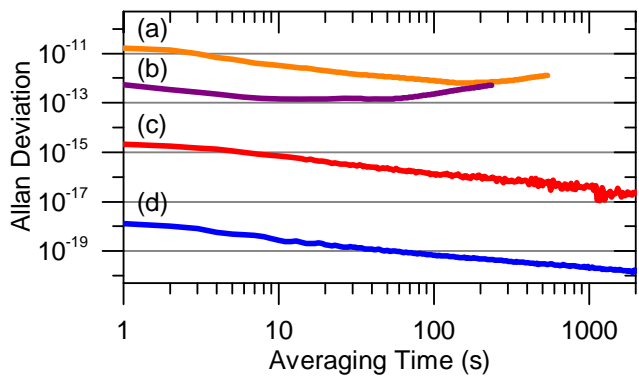

Fig. 3. Fractional frequency instability measurements of the LFC normalized by the detected optical frequency. Beat between the LFC and a 657 nm cavity-stabilized laser when the LFC is stabilized to a Rb clock (a) and a hydrogen maser (b). At short times, the instability is limited by the Rb clock or maser, while at long times the instability is dominated by the cavity drift (~0.4 Hz/sec.) Curves (c) and (d) are the instability of $f_{CEO}$ measured out-of-loop between 660 nm and 1340 nm and in-loop between 1 μm and 2 μm, respectively.

Additional independent evidence of the good coherence of the broad bandwidth continuum is provided by measurement of a beat note between the visible comb light and a cavity-stabilized 657 nm laser =[18]=[18] (linewidth ~ 1 Hz). A signal-to-noise ratio of >25 dB (300 kHz resolution bandwidth) was achieved and was measured using a frequency counter. The instability of this beat is shown in Fig. 3. When the mode-locked laser is referenced to a Rb microwave clock, the measured stability is consistent with the $2 \times 10^{-11}$ fractional stability of the Rb clock itself; referencing the laser to the H maser reduces the short-term instability by more than an order of magnitude. At time scales of tens of seconds, both measurements show the long-term drift of the optical cavity to which the CW laser is stabilized.

In summary, we have demonstrated an all-fiber source of high quality pulses at 1040 nm compressible to 73 fs. The short duration and high peak power of these pulses enable coherent and continuous extension of the Er:fiber-based frequency comb to visible wavelengths. We envision this source will be useful for broad-bandwidth spectroscopy and for generating, with filtering, a LFC-based calibration of high-precision astronomical spectrographs in the near-infrared and visible.

This work was supported by NIST and the NSF. We thank F. Adler and J.-Y. Kim for helpful comments, R. Fox for providing the stabilized 657 nm laser light, and M. Hirano and Sumitomo Electric Industries for providing the highly nonlinear fiber. This paper is a contribution from the US government and is not subject to copyright in the US. Mention of specific trade names is for technical information only and does not constitute an endorsement by NIST.

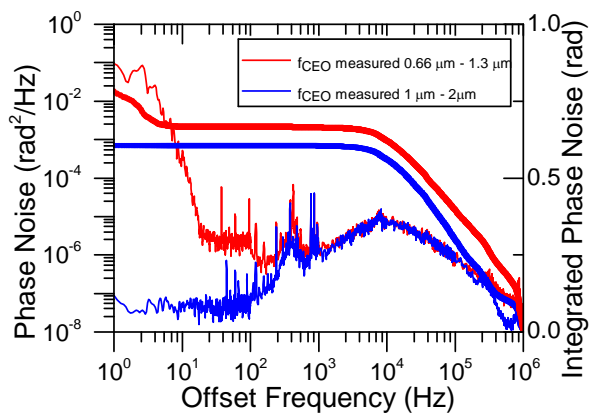

Fig. 4. Phase noise of $f_{CEO}$ measured on the signal used to lock the laser (blue) and the out-of-loop Yb branch (red).